\documentclass[11pt,amsmath,amssymb,]{revtex4}
\usepackage{amsfonts}
\usepackage{amsmath}
\usepackage{amssymb}
\usepackage{array}
\usepackage{bm}
\usepackage{braket}
\usepackage{breqn}
\usepackage{CJK}
\usepackage{color}
\usepackage{dcolumn}
\usepackage{epsfig}
\usepackage{epsf}
\usepackage{epstopdf}
\usepackage{float}
\usepackage{graphicx}
\usepackage{graphics}
\usepackage{harpoon}
\usepackage{indentfirst}
\usepackage{mathrsfs}
\usepackage{multirow}
\usepackage{xcolor}

\newcommand{\myvec}[1]  {\overset{\rightharpoonup}{\textbf{#1}}}
\newcommand {\sla}[1]{ #1 \!\!\!/}

\begin{document}

\title{Decay widthes of $^3 P_J$ charmonium to $DD,DD^*,D^*D^*$ and corresponding mass shifts of $^3 P_J$ charmonium}

\author{Hui-Yun Cao, Hai-Qing Zhou\protect\footnotemark[1]\protect\footnotetext[1]{E-mail: zhouhq@seu.edu.cn}\\School of Physics, Southeast University, Nanjing 211189, China\\}

\date{\today}

\begin{abstract}
In this work, we calculate the amplitudes of the processes $c\bar c({^3P_J}) \rightarrow DD,DD^*, D^*D^* \rightarrow c\bar c({^3P_J})$ in the leading order of the nonrelativistic expansion. The imaginary parts of the amplitudes are corresponding to the  branch decay widthes of the  charmonium $c\bar c({^3P_J}) \rightarrow DD,DD^*, D^*D^*$ and the real parts are corresponding to the mass shifts of the charmonium $c\bar c({^3P_J})$ due to these decay channels. After absorbing the polynomial contributions which are pure real and include the UV divergences, the ratios between the branch decay widthes and the corresponding mass shifts are only dependent on the center-of-mass energy.  We find the decay widthes and the mass shifts of the $^3P_2$ states are exact zero in the leading order. The ratios between the branch decay widthes and the mass shifts for the $^3P_0, {^3P_1}$ states are larger than 5 when the center-of-mass energy is above the $DD,DD^*, D^*D^*$ threshold. The dependence of the mass shifts on the center-of-mass energy is nontrivial especially when the center-of-mass energy is below the threshold. The analytic results can be extended to the $b$ quark sector directly.
\end{abstract}

\maketitle

\section{Introduction}

The energy spectrum of hadrons is a basic topic in the strong interaction. Up to now, it is still an unsolved problem due to the complex nonperturbative property of QCD. In literatures, many phenomenological models have been developed to study this problem in the quark level, such as the quark model \cite{qurk model}, QCD sum rules \cite{QCD sum rules}, Bethe-Salpeter equation \cite{DSE and BSE}, and etc. In these methods, usually the annihilation effects are neglected since they are much smaller than the non-perturbative potential. Physically, if the annihilation effect can be taken as small comparing with the interaction which binds the quarks, then the imaginary part of the annihilation amplitude is corresponding to the branch decay width and the real part is corresponding to the perturbative mass shift. Theoretically such annihilation effects should be considered and estimated carefully when aiming to understand the energy spectrum precisely.

Experimentally, since 2003 many new charmonium-like states are reported by the collaborations of Belle \cite{Belle}, CDF \cite{CDF}, D0 \cite{D0}, BABAR \cite{BaBar}, Cleo-C \cite{CLEO}, LHCb \cite{LHCb}, BES \cite{BESIII}, and CMS \cite{CMS}. These charmonium-like states cannot be well understood in the traditional quark model and their masses usually lie above the open charm threshold where some new decay modes are opened. In the previous study \cite{Zhou2019-TwoGluonAnihilation}, we studied the mass shifts of $^1S_0$ and $^3P_J$ heavy quarkonia due to the transition $q\bar{q}\rightarrow 2g\rightarrow q\bar{q}$. Physically, when the masses of the states lie above the threshold of $D$ or $D^*$ pairs, the transitions $c\bar c$ to these mesons' pairs are opened. It is natural that these opened channels not only result in the visible branch decay widthes but also give contributions to the mass shifts of the corresponding charmonium. When the masses of the charmonium lie about the threshold of the meson pairs, one can expect that the nonrelativistic expansion is available, which means that one can take the mesons $D,D^*$ like the heavy quark in the nonrelativistic QCD to construct the effective nonrelativistic interactions order by order. In this work, we follow this spirit to calculate the amplitudes of $c\bar c (^3P_J) \rightarrow DD, DD^*, D^*D^* \rightarrow c\bar c (^3P_J)$ in the leading order of non-relativistic expansion. The imaginary parts of the results are corresponding to the branch decay widthes which can be used to determine the effective coupling constants. Furthermore, if these annihilation interactions are much smaller than the binding interaction, then the real parts can be used to estimate the corresponding mass shifts.

We organize the paper as follow. In Sec. II we describe the basic frame to calculate the amplitudes of  $c\bar c (^3P_J) \rightarrow DD, DD^*,D^*D^* \rightarrow \bar c (^3P_J)$ in the leading order of nonrelativistic expansion, in Sec. III we give the analytic results for the amplitudes in the leading order of nonrelativistic expansion, in Sec. IV, we present some numerical results to show some properties in detail.

\section{Basic Formula}

When the mass of the charmonium is about $2m_D$ or $2m_{D^*}$ with $m_{D,D^*}$ being the masses of the $D,D^*$ mesons, the three-momenta of the $c$ quarks and the mesons in the decay channels $c\bar{c} (^3P_J)\rightarrow DD,DD^*,D^*D^*$ are much smaller than $c$ quarks' mass $m_c$ or $m_{D,D^*}$. In this case, one can take $m_c\approx m_D \approx m_{D^*}$ as the large scale comparing with $\Lambda_{\textrm{QCD}}$ and expand the interaction on the small variables $|\myvec{q}|/m_c$ with $\myvec{q}$ the three-momenta of the $c$ quarks and the mesons. This nonrelativistic expansion is similar with the spirit of NRQCD where the contact four point interactions are introduced. Different from NRQCD, now there is no hard gluon in the decay channels $c\bar{c} (^3P_J)\rightarrow DD,DD^*,D^*D^*$, but only nonrelativistic heavy quarks and heavy mesons. This means that there are only contact interactions between the $c$ quarks and the $D,D^*$ mesons. In the leading order of $|\myvec{q}|/m_c$, naively the most general interactions with $C,P,T$ invariance can be written as follows:
\begin{eqnarray}
{\cal L}_1&=&g_{a} \overline{\psi}\psi \phi_D \phi_D, \nonumber\\
{\cal L}_2&=&g_{b}\overline{\psi}\gamma^5 \gamma_{\mu}\psi \phi_D A^{\mu}_{D^*}+ \textit{h.c.},\nonumber\\
{\cal L}_3&=&g_{c} \overline{\psi}\psi A_{\mu D^*} A^{\mu}_{D^*},
\end{eqnarray}
where $\psi,\phi_D,A^{\mu}_{D^*}$ are the fields of the $c$ quark, the $D$ meson, and the $D^*$ meson, respectively. Here we do not assume that there is spin asymmetry between the $D$ and $D^*$ mesons since the dynamics of the light quarks insider the $D$ and $D^*$ mesons may break the spin symmetry strongly. This means that the couplings $g_{a,b,c}$ are independent.

\begin{figure*}[htbp]
\centering
\includegraphics[width=15cm]{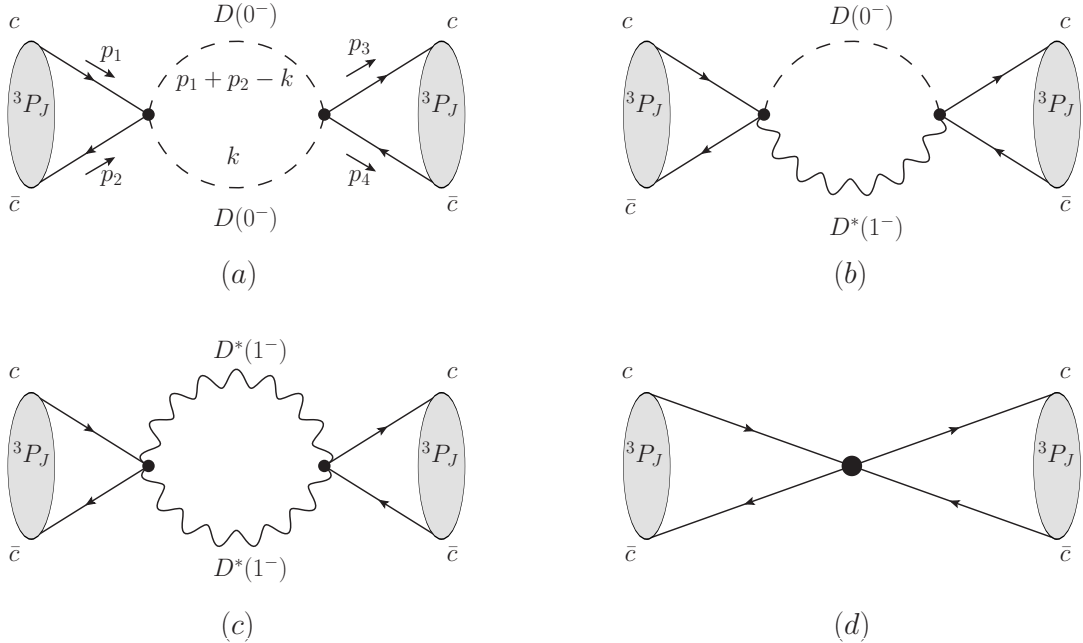}
\caption{The diagrams for $c\bar c(^3P_J)\rightarrow c\bar c(^3P_J)$ process where $(a,b,c,d)$ are corresponding to $c\bar c \rightarrow DD \rightarrow c\bar c$, $c\bar c \rightarrow DD^* \rightarrow c\bar c$, $c\bar c \rightarrow D^*D^* \rightarrow c\bar c$, and $c\bar c\rightarrow c\bar c$ via contact interactions.}
\label{figure:Feynman-diagram}
\end{figure*}

By these interactions, the Feynman diagrams for the amplitudes of $c\bar c ({^3P_J})\rightarrow DD, DD^*,D^*D^* \rightarrow c\bar c({^3P_J})$ in the leading order are showed in Fig. \ref{figure:Feynman-diagram}$(a,b,c)$.

Similar with any effective theory, usually the contract interactions are needed to absorb the UV divergence in the loop diagrams. To absorb the UV divergence in Fig. \ref{figure:Feynman-diagram}$(a,b,c)$, the following contact interactions are needed:
\begin{eqnarray}
{\cal L}^{c}_1&=&g_{10} [\overline{\psi}\psi] [\overline{\psi}\psi]-g_{11}\Big(\partial_\mu\partial^\mu [\overline{\psi}\psi]\Big) [\overline{\psi}\psi]+g_{12}\Big(\partial_\mu\partial^\mu\partial_\nu\partial^\nu[\overline{\psi}\psi]\Big)[\overline{\psi}\psi] , \nonumber\\
{\cal L}^{c}_2&=&g_{20}[\overline{\psi}\gamma^5 \gamma_{\mu}\psi][\overline{\psi}\gamma^5 \gamma^{\mu}\psi]- g_{21}\Big(\partial_\nu\partial^\nu[\overline{\psi}\gamma^5 \gamma_{\mu}\psi]\Big)[\overline{\psi}\gamma^5 \gamma^{\mu}\psi],
\end{eqnarray}
where the higher orders of the interactions are also kept. We want to point out that we just write down such contact interactions here to show the exact cancellation of the UV divergence and the polynomial contributions. In the practical calculation, one can get the same final results even without knowing the form of the contact interactions. The Feynmann diagram for the contribution due to these contact interactions is showed in Fig. \ref{figure:Feynman-diagram}$(d)$.

In the center of mass frame, we choose the four external momenta as follows:
\begin{eqnarray}
p_1&\overset{def}{=}&\frac{P}{2}+q_i,  \quad  p_2 \overset{def}{=} \frac{P}{2}-q_i, \nonumber\\
p_3&\overset{def}{=}&\frac{P}{2}+q_f,  \quad  p_4 \overset{def}{=} \frac{P}{2}-q_f.
\end{eqnarray}
For simplicity we define $P \overset{def}{=} (\sqrt{s},0,0,0)$ and use the instantaneous approximation for $q_{i,f}$ which means that we assume $q_i= (0,\textbf{q}_i)$ and  $ q_f= (0,\textbf{q}_f)$, where we use the bold formatting to refer to the three momentum here and in the following.

To project the $c\bar c$ pairs to the $^3P_J$ states  we use the project matrices in the on-shell case \cite{project-operator-1,project-operator-2,project-operator-3} which are defined as follows:
\begin{eqnarray}
\sum \bar{\nu}(\textbf{p}_2,s_2)T u(\textbf{p}_1,s_1)<\frac{1}{2}s_1;\frac{1}{2}s_2|1s_i> \overset{def}{=} \text{Tr}[T\Pi_i(s_i)], \nonumber\\
\sum \bar{u}(\textbf{p}_3,s_3)T\nu(\textbf{p}_4,s_4)<\frac{1}{2}s_3;\frac{1}{2}s_4|1s_f> \overset{def}{=} \text{Tr}[T\Pi_f(s_f)],
\end{eqnarray}
where the Clebsch-Gordan coefficients are the standard ones as in Ref. \cite{project-operator-2}, and the Dirac spinors are normalized as $u^+u=\nu^+\nu=1$, whose definitions are expressed as
\begin{eqnarray}
u(\textbf{p}_1,s_1) \overset{def}{=} \frac{\sla{\overline{p}}_1+m}{\sqrt{E_1(E_1+m)}}\left(
\begin{array}{c}
\xi^{s_1} \\
0 \\
\end{array}
\right),  \nonumber\\
\nu(\textbf{p}_2,s_2) \overset{def}{=} \frac{-\sla{\overline{p}}_2+m}{\sqrt{E_2(E_2+m)}}\left(
\begin{array}{c}
0 \\
\eta^{s_2} \\
\end{array}
\right),
\end{eqnarray}
with $E_{1,2} =\sqrt{|\textbf{p}_{1,2}|^2+m_c^2}$, $\overline{p}_{1,2}=(E_{1,2},\textbf{p}_{1,2})$, $\xi^{1/2}=(1,0)^T$, $\xi^{-1/2}=(0,1)^T$, $\eta^{1/2}=(0,1)^T$, and $\eta^{-1/2}=(-1,0)^T$. Finally the project matrices can be written as
\begin{eqnarray}
\Pi_i(s_i) &=&  N_{i} (\sla{\overline{p}}_1+m_c)(2E_i+\sla{\overline{p}}_1+\sla{\overline{p}}_2)\sla{\epsilon}_p(s_i)(-\sla{\overline{p}}_2+m_c),\nonumber\\
\Pi_f(s_f) &=&  N_{f}  (-\sla{\overline{p}}_4+m_c)\sla{\epsilon}_p^*(s_f)(2E_f+\sla{\overline{p}}_3+\sla{\overline{p}}_4)(\sla{\overline{p}}_3+m_c),
\end{eqnarray}
where $E_{i,f}= \sqrt{|\textbf{q}_{i,f}|^2+m_c^2}$, and
\begin{eqnarray}
\epsilon_p^{\mu}(0) &\overset{def}{=}& (0,0,0,1), \nonumber\\
\epsilon_p^{\mu}(\pm 1) &\overset{def}{=}& (0,\mp 1 ,-i,0)/\sqrt{2},
\end{eqnarray}
and $N_{i,f}$ are the normalized global factors which can be expressed as follows in the nonrelativistic limit
\begin{eqnarray}
N_{i,f}=-\frac{1}{8\sqrt{2}E_{i,f}^2(E_{i,f}+m_c)}.
\end{eqnarray}

In principle the form of the project matrix for a bounded $c\overline{c}$ pair should be deduced from the Bethe-Salpeter wave funciton or similar Lorentz covariant matrix element, while in the ultra nonrelativistic limit the above expressions are expected to be correct.

In the leading order of nonrelativistic expansion, the structure of a meson $H(^3P_J)$ can be expressed as follow:
\begin{eqnarray}
|H(^3P_J)\rangle  \sim \phi(|\textbf{p}|) \frac{\delta_{ij}}{\sqrt{N_c}}|q^i\bar{q}^j(^3P_J)\rangle,
\end{eqnarray}
where $N_c=3$ and $\phi(|\textbf{p}|)$ is the wave function of $H(^3P_J)$ in the momentum space which is defined as
\begin{eqnarray}
\phi(|\textbf{p}|) Y_{1m}(\Omega_{\textbf{p}}) \overset{def}{=} \int\frac{d^3\textbf{r}}{(2\pi)^3}e^{-i\textbf{p}\cdot \textbf{r}}R_1(|\textbf{r}|)Y_{1m}(\Omega_{\textbf{r}}),
\end{eqnarray}
with the normalization condition
\textcolor{black}{
\begin{eqnarray}
\int d|\textbf{r}||\textbf{r}|^2R^2_1(|\textbf{r}|)=1.
\end{eqnarray}}

Combining the structure of $H(^3P_J)$ and the project matrices, the expression for the amplitudes in the leading order can be expressed as
\begin{eqnarray}
{\cal M}^{(X)}(^3P_J) &=& \int d |\textbf{q}_i| d |\textbf{q}_f| |\textbf{q}_i|^2  |\textbf{q}_f|^2 \phi(|\textbf{q}_f|)\phi^*(|\textbf{q}_i|)\overline{G}^{(X)}(^3P_J),
\end{eqnarray}
where the index $(X)$ refers to $(a,b,c,d)$ which are corresponding to the contributions from the diagrams $(a)$, $(b)$, $(c)$ and $(d)$ showed in Fig. \ref{figure:Feynman-diagram}, respectively. $\overline{G}^{(X)}(^3P_J)$ are expressed as
\begin{eqnarray}
\overline{G}^{(X)}(^3P_J) &=& \sum_{s_i,s_f}<JJ_z|1s_f;1m_f> <JJ_z|1s_i;1m_i> \int d\Omega_{\textbf{q}_i} d\Omega_{\textbf{q}_f} Y_{1m_i}(\Omega_{\textbf{q}_i}) \nonumber\\
&& \times \  Y_{1m_f}^*(\Omega_{\textbf{q}_f}) G^{(X)}(s_i,s_f),
\end{eqnarray}
with
\begin{eqnarray}
G^{(a)}(s_i,s_f)  &=& -ic_f \mu^{2\epsilon}\int \frac{d^d k}{(2\pi)^d} \text{Tr}[T_1  \Pi_{i}(s_i)] \text{Tr}[T_1 \Pi_{f}(s_f)] S(k) S(p_1+p_2-k),  \nonumber\\
G^{(b)}(s_i,s_f)  &=& -ic_f \mu^{2\epsilon}\int \frac{d^d k}{(2\pi)^d} \text{Tr}\big[T_2^{\mu}  \Pi_{i}(s_i)\big] \text{Tr}\big[T_2^{\nu} \Pi_{f}(s_f)\big] D_{\mu \nu}(k) S(p_1+p_2-k), \nonumber\\
G^{(c)}(s_i,s_f)  &=& -ic_f \mu^{2\epsilon}\int \frac{d^d k}{(2\pi)^d} \text{Tr}[T_3^{\mu\rho} \Pi_{i}(s_i)] \text{Tr}[T_3^{\nu\omega} \Pi_{f}(s_f)] D_{\mu \nu}(k) D_{\rho \omega}(p_1+p_2-k),\nonumber \\
G^{(d)}(s_i,s_f)  &=&  -ic_f \mu^{2\epsilon}\Big(\text{Tr}[T_4\Pi_{i}(s_i) ] \text{Tr}[\Pi_{f}(s_f)] + \text{Tr}[T_5^{\mu} \Pi_{i}(s_i)] \text{Tr}[\gamma_{5}\gamma_\mu  \Pi_{f}(s_f)] \Big),
\end{eqnarray}
where $d=4-2\epsilon$ is the dimension, $\mu$ is the introduced energy scale, $c_f=\frac{\delta_{ij}}{\sqrt{N_c}}\delta_{ij}\frac{\delta_{i'j'}}{\sqrt{N_c}}\delta_{i'j'}=3$ is the color factor, and
\begin{eqnarray}
T_1  &=& ig_a, \nonumber\\
T_2^{\mu}&=&ig_b \gamma^5 \gamma^{\mu},\nonumber\\
T_3^{\mu\rho}  &=& ig_c g^{\mu\rho},\nonumber\\
T_4  &=& i(g_{10}+g_{11}s+g_{12}s^2),\nonumber\\
T_5^{\mu}  &=& i(g_{20} +g_{21}s)\gamma_5 \gamma^{\mu},
\end{eqnarray}
and the propagators of the pseudoscalar $S$ and the vector $D_{\mu \nu}$ are defined as
\begin{eqnarray}
S(k) &=& \frac{i}{k^2-m_D^2+i \varepsilon}, \nonumber\\
D_{\mu \nu}(k) &=& \frac{-i(g^{\mu \nu}-  \frac{k^{\mu}k^{\nu}}{m_{D^*}^2})}{k^2-m_{D^*}^2+i \varepsilon}.
\end{eqnarray}

In the practical calculation, the package FeynCalc \cite{FeynCalc} is used to do the trace in the $d$ dimension. The packages FIESTA \cite{FIESTA} and PackageX \cite{PackageX} are independently used to do the loop integration for double check. After the loop integrations, $G^{(X)}(s_i,s_f)$ can be expressed in the following form:
\begin{eqnarray}
G^{(X)}(s_i,s_f) &=& C_1^{(X)} \epsilon_p(s_i)\cdot \epsilon_p^*(s_f) + C_2^{(X)}\epsilon_p(s_i)\cdot \hat{q}_i \ \epsilon_p^*(s_f)\cdot \hat{q}_f + C_3^{(X)}\epsilon_p(s_i)\cdot \hat{q}_f \ \epsilon_p^*(s_f)\cdot \hat{q}_i,
\end{eqnarray}
where $C^{(X)}_i$ can be expressed as
\begin{eqnarray}
C^{(X)}_i = \sum_{n=0}^1 C^{(X)}_{in}(|\textbf{q}_i|,|\textbf{q}_f|)(\hat{q_i}\cdot \hat{q}_f)^n,
\end{eqnarray}
with $\hat{q}_{i,f}\overset{def}{=}q_{i,f}/|\textbf{q}_{i,f}|$, respectively.

To get the coefficients $\overline{G}^{(X)}(^3P_J)$, usually the sums of the spins and the integrations of the angles are calculated independently to simplify the expressions \cite{spin-angle-independently}. In our calculation, we directly calculate the sums of the spins and the integrations of the angles together after getting the expressions of $C^{(X)}_{in}$. This method is more efficient and has been used in our previous work \cite{Zhou2019-TwoGluonAnihilation}. The relevant expressions are listed in the Appendix.

\section{The energy shift of $^3P_J$ in the leading order}
We expand $\overline{G}^{(X)}(^3P_J)$ on $|\textbf{q}_{i}|,|\textbf{q}_{f}|$ to order 1  as following forms:
\begin{eqnarray}
\overline{G}^{(a,b,c)}(^3P_J) &=& 3g^2_{a,b,c}N_{i}N_{f} m_c^4\Big[ |\textbf{q}_{i}||\textbf{q}_{f}|c^{(a,b,c)}_{J} + \textrm{higher order} \Big],\nonumber\\
\overline{G}^{(d)}(^3P_J) &=& 3N_{i}N_{f} m_c^4\Big[ |\textbf{q}_{i}||\textbf{q}_{f}|c^{(d)}_{J} + \textrm{higher order} \Big].
\label{equation:coe-initial}
\end{eqnarray}

Here we want to emphasis that the contributions $\overline{G}^{(d)}(^3P_J)$ are used to absorb the UV divergences in $\overline{G}^{(a+b+c)}(^3P_J)$ and give no contributions to the decay widthes of $^3P_J$ states. The finite parts of the contributions $\overline{G}^{(d)}(^3P_J)$ are arbitrary. Actually, they not only absorb the UV divergences but also absorb the polynomial contributions in $\overline{G}^{(a+b+c)}(^3P_J)$. This situation is a little different from the results in the $c\bar c(^3P_J)\rightarrow 2g \rightarrow c\bar c(^3P_J) $ cases where there are no any contact interactions in the original QCD interaction. The important point is that these absorptions are universal and independent on the processes, and we discuss the details in the following subsection.




\subsection{The energy shift of $^3P_0$ state}
After the loop integration, the sum of the spins, the integration of the angles, and the Taylor expansion,  we get the following results in the $^3P_0$ channel.
\begin{eqnarray}
c^{(a)}_0 &=& c^{(a)}_{0,poly}+\frac{256\sqrt{s(s-4m_D^2)}}{\pi s}\ln\big[\frac{2m_D^2-s+\sqrt{s(s-4m_D^2)}}{2m_D^2}+i\varepsilon \big], \nonumber \\
c^{(b)}_0 &=&0, \nonumber \\
c^{(c)}_0 &=&c^{(c)}_{0,poly}+\frac{64[(s-2m_{D^*}^2)^2+8m_{D^*}^4]\sqrt{s(s-4m_{D^*}^2)}}{\pi s m_{D^*}^4} \ln\big[\frac{2m_{D^*}^2-s+\sqrt{s(s-4m_{D^*}^2)}}{2m_{D^*}^2}+i\varepsilon \big],\nonumber \\
c^{(d)}_{0}&=&c^{(d)}_{0,poly},
\label{equation:c0-abcd}
\end{eqnarray}
where $c^{(a,c,d)}_{0,poly}$ are some polynomial functions on $s$ which include the UV divergences and are expressed as follows:
\begin{eqnarray}
c^{(a)}_{0,poly} &=& \frac{256}{\pi}(2+\frac{1}{\overline{\epsilon}_{\textrm{UV}}}+\ln\frac{\mu_{\textrm{UV}}^2}{m_D^2}), \nonumber \\
c^{(c)}_{0,poly} &=& \frac{64}{\pi m_{D^*}^4}\Big[4(4+\frac{3}{\overline{\epsilon}_{\textrm{UV}}}+3\ln\frac{\mu_{\textrm{UV}}^2}{m_{D^*}^2})m_{D^*}^4
-2(5+\frac{3}{\overline{\epsilon}_{\textrm{UV}}}+3\ln\frac{\mu_{\textrm{UV}}^2}{m_{D^*}^2})m_{D^*}^2s\nonumber\\
&&~~~~~~~~~~~+(2+\frac{1}{\overline{\epsilon}_{\textrm{UV}}}+\ln\frac{\mu_{\textrm{UV}}^2}{m_{D^*}^2})s^2\Big],\nonumber\\
c^{(d)}_{0,poly} &=& \frac{256}{\pi^3} (g_{10}+g_{11}s+g_{12}s^2),
\label{equation:c0-abc-pily}
\end{eqnarray}
with$\frac{1}{\overline{\epsilon}_{\textrm{UV}}}=\frac{1}{\epsilon_{\textrm{UV}}}-\gamma_E+\log(4\pi)$.

An important property of the two contributions $c^{(a,c)}_{0,poly}$ is that they can be absorbed by the contact interactions ${\cal L}_1^{c}$ independently. These contact interactions are independent and give no contributions to the decay widthes of the charmonium. This means that their effects can be absorbed by the models which are used to calculate the energy spectrum  and do not include the annihilation effects. Here we are only interested in the mass shifts due to the decay modes, then we only focus on the contributions including the imaginary parts due to the loop calculation and neglect the terms $c^{(a,c)}_{0,poly}$. The choices of $g_{10,11,12}$ which can cancel all the  polynomial contributions in $c_{0}^{(a,c)}$ can be got directly.

From Eq. (\ref{equation:c0-abcd}), one can easily get the imaginary parts as follows:
\begin{eqnarray}
\text{Im}[c^{(a)}_0] &=& \frac{256\sqrt{s(s-4m_D^2)}}{s}  \theta(s-4m_D^2), \nonumber \\
\text{Im}[c^{(b)}_0] &=&0, \nonumber\\
\text{Im}[c^{(c)}_0] &=&\frac{64[(s-2m_{D^*}^2)^2+8m_{D^*}^4]\sqrt{s(s-4m_{D^*}^2)}}{s m_{D^*}^4}\theta(s-4m_{D^*}^2), \nonumber\\
\text{Im}[c^{(d)}_0] &=&0.
\end{eqnarray}

Matching the amplitude with the corresponding amplitude in quantum mechanism with a perturbativel  potential, one has
\begin{eqnarray}
{\cal M}(^3P_J)= -\langle H(^3P_J)|V_{eff}|H(^3P_J)\rangle.
\end{eqnarray}

Finally the decay widthes of $^3P_0$ to $DD$ and $D^*D^*$ in the leading order are expressed as follows:
\begin{eqnarray}
\Gamma(^3P_0\rightarrow DD ) &=& 2 \text{Im}[{\cal M}^{(a)}(^3P_0)] = \textcolor{black}{\frac{27g_a^2}{8\pi^2}}N_iN_fm_c^4 \text{Im}[c_{0}^{(a)}]|R_1^{(1)}(0)|^2, \nonumber\\
\Gamma(^3P_0\rightarrow DD^*) &=& 2 \text{Im}[{\cal M}^{(b)}(^3P_0)] =0, \nonumber\\
\Gamma(^3P_0\rightarrow D^*D^* ) &=& 2 \text{Im}[{\cal M}^{(c)}(^3P_0)] =  \textcolor{black}{\frac{27g_c^2}{8\pi^2}}N_iN_fm_c^4 \text{Im}[c_{0}^{(c)}]|R_1^{(1)}(0)|^2,
\end{eqnarray}
where we have used the relation
\begin{eqnarray}
\int \phi(p) p^{2n+3} dp = (-1)^n \frac{2n+3}{4\pi} R_1^{(2n+1)}(|\textbf{r}|)\Big|_{|\textbf{r}|=0}.
\end{eqnarray}

The corresponding mass shifts labeled as $\Delta m(^3P_0)$ are expressed as
\begin{eqnarray}
\Delta m(^3P_0) &=& -\text{Re}[{\cal M}^{(a+b+c)}(^3P_0)]  \nonumber\\
&=&-\frac{\text{Re}[\overline{c}_{0}^{(a)}]}{2\text{Im}[c_{0}^{(a)}]}\Gamma(^3P_0\rightarrow DD)-\frac{\text{Re}[\overline{c}_{0}^{(c)}]}{2\text{Im}[c_{0}^{(c)}]}\Gamma(^3P_0\rightarrow D^*D^*),
\end{eqnarray}
where $\overline{c}_{0}^{(a,c)}=c_{0}^{(a,c)}-c_{0,poly}^{(a,c)}$.

\subsection{The energy shift of $^3P_1$ state}
In the $^3P_1$ channel, we have the following results
\begin{eqnarray}
c^{(a)}_1 &=&0, \nonumber \\
c^{(b)}_1 &=&c^{(b)}_{1,poly}+\frac{128}{9\pi s^2m_{D*}^2}A(A^2+12sm_{D*}^2)\ln\Big[\frac{A-s+m_D^2+m_{D^*}^2}{2m_Dm_{D^*}}+i\varepsilon \Big] , \nonumber \\
c^{(c)}_1 &=&0, \nonumber\\
c^{(d)}_1 &=&c^{(d)}_{0,poly},
\label{equation:c1-abc}
\end{eqnarray}
with
\begin{eqnarray}
A&=& \sqrt{\big[ s-(m_D-m_{D*})^2\big] \big[s-(m_D+m_{D*})^2\big]}.
\end{eqnarray}

The polynomial terms are expressed as
\begin{eqnarray}
c^{(b)}_{1,poly} &=& \sum\limits_{n=-2}^{1}s^nc_{1;n}^{(b)},\nonumber\\
c^{(d)}_{1,poly} &=& -\frac{512}{3\pi^3}(g_{20}+g_{21}s),
\label{equation:c1-b}
\end{eqnarray}
with
\begin{eqnarray}
c_{1;-2}^{(b)}&=&\frac{64}{9\pi m_{D^*}^2}(m_{D^*}^2-m_D^2)^3\ln\frac{m_D^2}{m_{D^*}^2},\nonumber\\
c_{1;-1}^{(b)}&=&\frac{64}{9\pi m_{D^*}^2}(m_{D^*}^2-m_D^2)\Big[2(m_{D^*}^2-m_D^2)+3(3m_{D^*}^2-m_D^2)\ln\frac{m_D^2}{m_{D^*}^2}\Big],\nonumber\\
c_{1;0}^{(b)}&=&-\frac{64}{9\pi m_{D^*}^2}\big[(2+\frac{6}{\overline{\epsilon}_{\textrm{UV}}}+6\ln\frac{\mu^2}{m_D^2})m_D^2
-2(23+\frac{9}{\overline{\epsilon}_{\textrm{UV}}}+9\ln\frac{\mu^2}{m_{D^*}^2})m_{D^*}^2+3(m_D^2-3m_{D^*}^2)\ln\frac{m_D^2}{m_{D^*}^2}\big],\nonumber\\
c_{1;1}^{(b)}&=&\frac{64}{27\pi m_{D^*}^2}(4+\frac{6}{\overline{\epsilon}_{\textrm{UV}}}+6\ln\frac{\mu^2}{m_D^2}+3\ln\frac{m_D^2}{m_{D^*}^2}).
\end{eqnarray}

At first glance, this property is very different from that in the $^3P_0$ channel due to the nonzero values of $c_{1,-2}$ and $c_{1,-1}$ which seems is un-physical. While actually when taking the nonrelativistic approximation $m_D\approx m_{D^*}$, one has $c_{1;-2},c_{1;-2} \approx 0$, this means that there contributions are very small in the nonrelativistic approximation and can be neglected. The numerical calculations also shows such property and we neglect these two terms.

Similarly,  the term $c^{(b)}_{1,poly}$ can be neglected when aiming to discuss the contributions from the annihilation effects.
The imaginary part of $c_1^{(b)} $ can be expressed as
\begin{eqnarray}
\text{Im}[c_1^{(a)}] &=&0,\nonumber\\
\text{Im}[c_1^{(b)}] &=&\frac{128}{9 s^2m_{D*}^2}A(A^2+12sm_{D*}^2) \theta\big(s-(m_D+m_{D^*}^2)\big),\nonumber\\
\text{Im}[c_1^{(c)}] &=&0,\nonumber\\
\text{Im}[c_1^{(d)}] &=&0.
\end{eqnarray}

In the leading order, the decay width of $^3P_1$ to $D D^*$, are expressed as
\begin{eqnarray}
\Gamma(^3P_1\rightarrow DD ) &=& 2 \text{Im}[{\cal M}^{(a)}(^3P_1)] = 0, \nonumber\\
\Gamma(^3P_1\rightarrow DD^* ) &=& 2 \text{Im}[{\cal M}^{(b)}(^3P_1)] =  \textcolor{black}{\frac{27g_b^2}{8\pi^2}}N_iN_fm_c^4 \text{Im}[c_{1}^{(b)}]|R_1^{(1)}(0)|^2,\nonumber\\
\Gamma(^3P_1\rightarrow D^*D^* ) &=& 2 \text{Im}[{\cal M}^{(c)}(^3P_1)] = 0,
\end{eqnarray}
and the corresponding mass shift labeled as $\Delta m(^3P_1)$ is expressed as
\begin{eqnarray}
\Delta m(^3P_1) &=& -\text{Re}[{\cal M}^{(a+b+c)}(^3P_1)] = -\frac{\text{Re}[\overline{c}_{1}^{(b)}]}{2\text{Im}[c_{1}^{(b)}]}\Gamma(^3P_1\rightarrow DD^*),
\end{eqnarray}
where $\overline{c}_{1}^{(b)}=c_{1}^{(b)}-c_{1,poly}^{(b)}$.

\subsection{The energy shift of $^3P_2$ state}

For $^3P_2$ state, we get
\begin{eqnarray}
c^{(a,b,c,d)}_2 &=&0.
\label{equation:c2-abc}
\end{eqnarray}

These results means that the decay widthes $\Gamma(^3P_2\rightarrow DD,DD^*,D^*D^* )$ are exact zero and there are no mass shifts for $^3P_2$ states in the leading order. This result is a strong property which can be tested by the experiments and be used to judge whether a state is pure $^3P_2$ heavy quarkonium or not.

Comparing our results with those results given by the $^3P_0$ model in Ref. \cite{Barnes2008}, one can find that both the two methods give the zero results for $c\overline{c}(^3P_0)\rightarrow DD$ and $c\overline{c}(^3P_1)\rightarrow DD^*$. But in Ref. \cite{Barnes2008}, the contributions $c\overline{c}(^3P_2)\rightarrow DD,DD^*, D^*D^*$ are nonzero and in the same order with the contributions in $c\overline{c}(^3P_1)\rightarrow DD^*,D^*D^*$. This property is very different from our results.  The calculation in Ref. \cite{Barnes2008} is based on the $^3P_0$ model where a light quark pair is dynamically produced in the vacuum  and the nonrelativistic wave functions of mesons are used to estimate the contributions. While our calculation is based on the general model independent interactions under the nonrelativistic expansion and the results are almost model independent except the approximation of the nonrelativistic expansion. In our calculation, all the dynamics of the light quark and $D,D^*$ meason is absorbed by the coupling constants in the leading order of the nonrelativistic expansion. On another hand, we only consider the contributions due to the annihilation effects and neglect the polynomial contribution since the latter is uncertain.

\section{Numerical results and discussion}

To show the properties of the above analytic results more clearly, we present some numerical results in this section. Firstly we want to emphasize that the absolute values of $\text{Re}[\overline{c}^{(a,b,c)}_{J}]$ and $\text{Im}[c^{(a,b,c)}_{J}]$ can not determine the physical decay widthes and the mass shifts directly, since there are global unknown constant factors. But the ratios of the mass shifts and the decay widthes $-\text{Re}[\overline{c}^{(a,b,c)}_{J}]/2\text{Im}[c^{(a,b,c)}_{J}]$ are model independent. This means that if the decay widthes are measured experimentally, the corresponding corrections to the masses of the heavy quarkonia can be got directly.

\begin{figure*}[htbp]
\centering
\includegraphics[width=16cm]{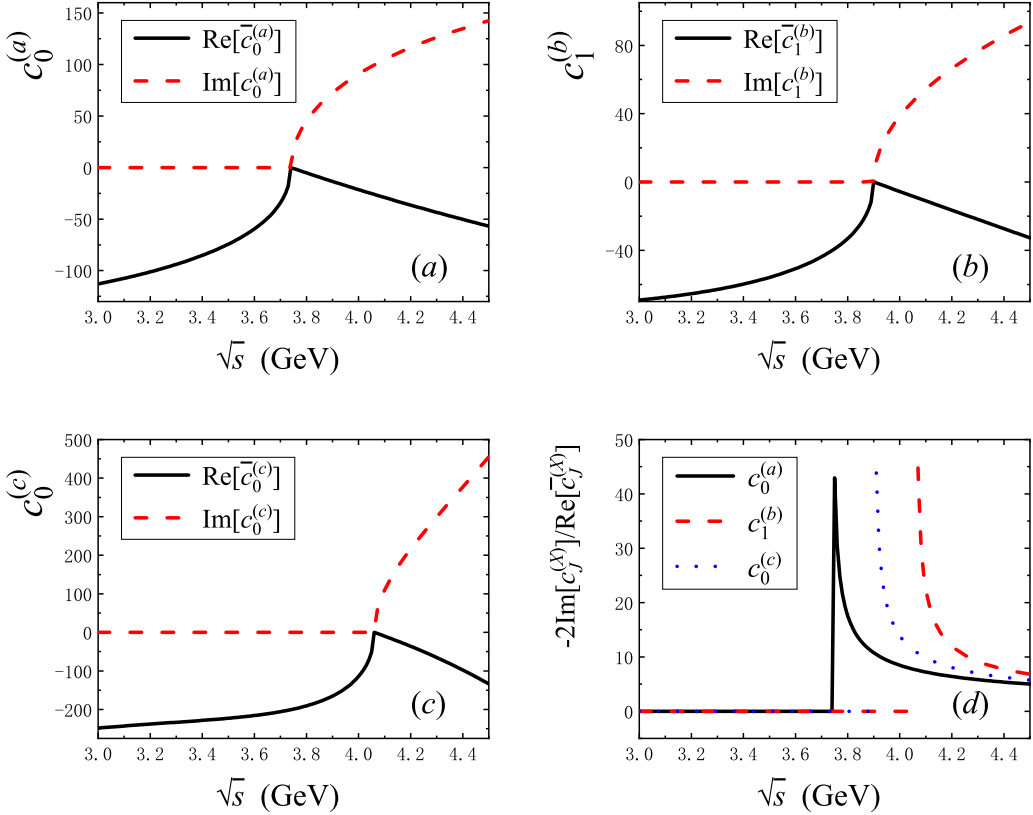}
\caption{Numerical results for $\text{Im}[c^{(a,b,c)}_{J}]$~~{\it vs.}~~$\sqrt{s}$, $\text{Re}[\overline{c}^{(a,b,c)}_{J}]$~~{\it vs.}~~$\sqrt{s}$ and  $-2\text{Im}[c^{(a,b,c)}_{J}]/\text{Re}[\overline{c}^{(a,b,c)}_{J}]$~~{\it vs.}~~$\sqrt{s}$. The sub figures $(a,b,c)$ are corresponding to $\text{Im}[c^{(a,b,c)}_{J}]$ and $\text{Re}[\overline{c}^{(a,b,c)}_{J}]$~~{\it vs.}~~$\sqrt{s}$, respectively.  The sub figure $(d)$ shows  the results for $-2\text{Im}[c^{(X)}_{J}]/\text{Re}[\overline{c}^{(X)}_{J}]$~~{\it vs.}~~$\sqrt{s}$.}
\label{figure:cXJ-sqrts}
\end{figure*}

In Fig. \ref{figure:cXJ-sqrts} the dependence of $\text{Im}[c^{(a,b,c)}_{J}]$, $\text{Re}[\overline{c}^{(a,b,c)}_{J}]$ and their ratios on $\sqrt{s}$ are presented, respectively, where we take $m_D=1.87\text{ GeV}$ and $m_{D^*}=2.01\text{ GeV}$ as inputs.

The numerical results presented in Fig. \ref{figure:cXJ-sqrts} show four interesting properties:

(1) The real parts $\text{Re}[\overline{c}^{(a,c)}_{0}]$ and $\text{Re}[\overline{c}^{(b)}_{1}]$ which are represented by the solid black curves are always negative. This means that after considering the annihilation effects, the masses of the $^3P_{0,1}$ states move up and the masses of $^3P_2$ states do not move.

(2) When $\sqrt{s}$ is on the threshold of $DD,DD^*$ or $D^*D^*$ the corresponding mass shifts are exact zero.

(3) When $\sqrt{s}$ is above the threshold, the mass shifts are much smaller than the corresponding decay widthes, the largest mass shift is about 1/5 of the corresponding decay width when $\sqrt{s}\approx 4.5$ GeV which is much larger than the threshold. This property gives a strong constrain on the mass shifts to all the $^3P_{0,1}$ states.

(4) When $\sqrt{s}$ is below the mass-shell, although the decay widthes are exact zero, but the mass-shifts are still nonzero and the dependence of $\text{Re}[\overline{c}^{(a,b,c)}_{J}]$ {\it vs.} $\sqrt{s}$ shows non-trivial property.

\begin{figure*}[htbp]
\centering
\includegraphics[width=12cm]{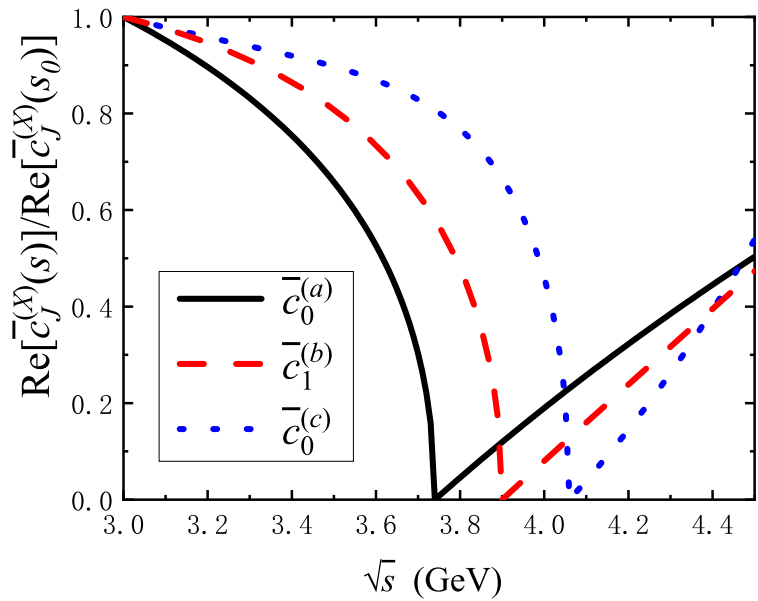}
\caption{Numerical results for the dependence of $\text{Re}[\overline{c}^{(a,b,c)}_{J}(s)]/\text{Re}[\overline{c}^{(a,b,c)}_{J}(s_0)]$~~{\it vs.}~~$\sqrt{s}$.}
\label{figure:ratio-cXJ-cXJs0}
\end{figure*}

To show the non-trivial dependence of $\text{Re}[\overline{c}^{(a,b,c)}_{J}]$ {\it vs.} $\sqrt{s}$ more clearly, we present the dependence of
$\text{Re}[\overline{c}^{(a,b,c)}_{J}(s)]/\text{Re}[\overline{c}^{(a,b,c)}_{J}(s_0)]|$ {\it vs.} $\sqrt{s}$ with $s_0=3$ GeV  in Fig. \ref{figure:ratio-cXJ-cXJs0}.  The curves in Fig. \ref{figure:ratio-cXJ-cXJs0} clearly show that when $\sqrt{s}$ increases from $3$ GeV to $4.5$ GeV the ratio of the mass shifts decreases from 1 to 0 at first and then increases from zero to 0.5.  For the states with the same quantum number, it means that the corresponding mass shifts are non-linear and can not be absorbed by some constants.

Experimentally, up to now there are still no definite results for the branch decay widthes $\Gamma(^3P_{0,1}\rightarrow DD,DD^*, D^*D^*)$\cite{Experiments}, this makes it difficult to determine the mass shifts certainly. The experiments reported that the decay widthes $\Gamma(X(3915),\chi_{c2}(3930)\rightarrow DD,DD^*, D^*D^*)$ are seen. By our calculation, we expect that the decay widthes $\Gamma(^3P_{2}\rightarrow DD,DD^*, D^*D^*)$ are zero in the leading order which suggests that the decay widthes $\Gamma(^3P_{2}\rightarrow DD,DD^*, D^*D^*)$ should be much smaller than $\Gamma(^3P_{0}\rightarrow DD, D^*D^*)$ and $\Gamma(^3P_{1}\rightarrow DD^*)$. A relative larger decay widthes of a resonance to $DD,DD^*, D^*D^*$ suggest that it maybe is not a pure $c\bar{c}(^3P_J)$ state. These properties are more reliable in the $b$ quark part and can be tested by the further precise experiments. Furthermore, the similar discussion can be extended to the $S$ wave states and compared with the similar studies in Ref. \cite{OSet-S-wave}. 

In summary, the nonrelativistic asymptotic behavior of the transitions $c\bar{c}(^3P_J) \rightarrow DD,DD^*,D^*D^* \rightarrow c\bar{c}(^3P_J)$ with $J=0, 1, 2$ are discussed.  We find that the decay widthes $\Gamma(^3P_{0}\rightarrow DD^*),\Gamma(^3P_{1}\rightarrow DD,D^*D^*)$ and $\Gamma(^3P_{2}\rightarrow DD,DD^*,D^*D^*)$ are exact zero in the leading order of nonrelativistic expansion.  For other channels, the ratios between the branch decay widthes and the mass shifts are larger than 5 when the center-of-mass energy is above the threshold. When below the threshold, the mass shifts are dependent on the center-of-mass energy nontrivially and can not be absorbed by a constant.

\section{Acknowledgements}
The author Hai-Qing Zhou would like to thank Zhi-Yong Zhou and Dian-Yong Chen for their helpful discussion.
This work was supported by the National Natural Science Foundation of China (Grand No. 11375044 and 11975075).
Hui-Yun Cao was supported by the Scientific Research Foundation of Graduate School of Southeast University (Grants No. YBPY1970).

\section{Appendix: The FIESTA integrations}
We define the following functions to refer to the results after summing the spins and integrating the angles:
\begin{eqnarray}
P(J,X,n) &\overset{def}=& \sum_{s_i,s_f}<JJ_Z|1s_f;1m_f> <JJ_Z|1s_i;1m_i> \int d\Omega_{\textbf{q}_i} d\Omega_{\textbf{q}_f} Y_{1m_i}(\Omega_{\textbf{q}_i}) Y_{1m_f}^*(\Omega_{\textbf{q}_f})(\hat q_i\cdot \hat q_f)^n X, \nonumber\\
\end{eqnarray}
where X are some functions dependent on $\hat q_{i}, \hat q_{f}, \epsilon_p(s_i)$, and $\epsilon_p^*(s_f)$ with $\hat q_{i,f} \overset{def}{=} q_{i,f}/|\textbf{q}_{i,f}|$, $P(J,X,n)$ are not dependent on $J_z$. When $J=0,1,2$ and $ n=0,1$, we have
\begin{eqnarray}
P(J,\epsilon_p(s_i)\cdot \epsilon_p^*(s_f),1) &=& \frac{4\pi}{3},\quad P(0,\epsilon_p(s_i)\cdot \hat{q_i} \ \epsilon_p^*(s_f)\cdot \hat{q_f},0) = 4\pi,  \nonumber\\
P(0,\epsilon_p(s_i)\cdot \hat{q_f} \ \epsilon_p^*(s_f)\cdot \hat{q_i},0) &=& \frac{4\pi}{3},\quad P(1,\epsilon_p(s_i)\cdot \hat{q_f} \ \epsilon_p^*(s_f)\cdot \hat{q_i},0) = -\frac{4\pi}{3}, \nonumber\\
P(2,\epsilon(s_i)\cdot \hat{q_f} \ \epsilon^*(s_f)\cdot \hat{q_i},0) &=& \frac{4\pi}{3},
\end{eqnarray}
and others are zero.

\end{document}